# TOP: Time-to-Event Bayesian Optimal Phase II Trial Design for Cancer Immunotherapy

Ruitao Lin[1], Robert L Coleman[2] and Ying Yuan[1]

[1]Department of Biostatistics, [2]Department of Gynecologic Oncology and Reproductive Medicine,

The University of Texas MD Anderson Cancer Center

## Abstract

**Purpose**

Immunotherapies have revolutionized cancer treatment. Unlike chemotherapies, immune agents often take longer time to show benefit, and the complex and unique mechanism of action of these agents renders the use of multiple endpoints more appropriate in some trials. These new features of immunotherapy make conventional phase II trial designs, which assume a single binary endpoint that is quickly ascertainable, inefficient and dysfunctional.

**Methods**

We propose a flexible and efficient time-to-event Bayesian optimal phase II (TOP) design. The TOP design is efficient in that it allows real-time "go/no-go" interim decision making in the presence of late-onset responses by using all available data, and maximizes the statistical power for detecting effective treatments. TOP is flexible in the number of interim looks and capable of handling simple and complicated endpoints under a unified framework. We conduct simulation studies to evaluate the operating characteristics of the TOP design.

**Results**

Compared to some existing designs, the TOP design shortens the trial duration and has higher power to detect effective treatment with well controlled type I errors.

**Conclusion**

The TOP design allows for making real-time "go/no-go" interim decisions in the presence of late-onset responses, and is capable of handling various types of endpoints under a unified framework. It is transparent and easy to implement as its decision rules can be tabulated and included in the protocol prior to the conduct of the trial. The TOP design provides a flexible, efficient and easy-to-implement method to accelerate and improve the development of immunotherapies.

## 1. Introduction

The primary objective of most phase II clinical trials is to evaluate the preliminary therapeutic effect of a new treatment, make "go/no-go" decisions on whether the new treatment warrants further investigation in a large-scale randomized phase III trial, and in some cases, provide justification for regulatory actions, such as attainment of breakthrough designation or accelerated FDA approval[1-2]. Conventional phase II clinical trial designs, e.g., Simon's two-stage design[3] and its variations[4-5], were developed for cytotoxic chemotherapy. These designs assume a single binary endpoint that is quickly ascertainable, e.g., objective response (OR) scored using Response Evaluation Criteria in Solid Tumors (RECIST)[6], with one or two prespecified interim analyses. If the interim analysis shows sufficiently acceptable toxicity and high response, the trial continues, otherwise is stopped for futility. Such a futility monitoring procedure serves as a critical component of modern clinical trials[7].

Immunotherapy has emerged as one of the most promising therapeutic modalities for the treatment of many cancers. Immuno-oncology agents differ from conventional cytotoxic agents in that they stimulate an innate immune response against the tumor. The novel mechanism of action of these drugs can impute clinical benefit in measures of response, ranging from traditional objective tumor size reduction to transient pseudo-progression followed by long-lasting partial response or stable disease[8]. As a result, the immune agents often take a longer time, e.g., a few months, to show benefit, compared to standard chemotherapeutic agents[9]. The immune-related RECIST (irRECIST)[10] and iRECIST[11] provide guidance to account for these unique tumor response characteristics by requiring CT confirmation of disease progression 4-6 weeks later. In addition, to better characterize treatment effect of immunotherapy, some immunotherapy trials adopt OR and progression free survival (PFS) as co-primary endpoints, in which the attainment of either or both endpoints would indicate the agent is promising.

These new features of immunotherapy have made conventional phase II trial designs inefficient and dysfunctional. Korn and Freidlin[12] showed that the delayed treatment effect causes substantial power loss for conventional designs. In addition, as anti-tumor response to immunotherapy requires a long assessment window to be scored, major logistical difficulties arise. For instance, suppose a trial defines a response assessment that takes up to 90 days to be scored. With an accrual rate of 3 patients/month and a planned interim analysis after 12 patients are accrued, on average only 6 patients will have completed the response assessment and the remaining 6 patients are still being followed-up with pending response outcomes. Without knowing the outcomes of the pending patients, the interim "go/no-go" decision often cannot be made under conventional phase II designs, e.g., Simon's two-stage design. One possible approach is to continue the accrual and postpone the interim analysis until all pending patients complete their response assessment. This over-run approach, however, may treat extra patients on potentially ineffective treatment and invalidates the operating characteristics of the design. The latter assumes that interim decisions are made in real time. The other approach is to suspend the accrual and wait until all patient's responses are scored before making "go/no-go" decisions. This, however, prolongs trial duration and delays treatment of new patients. These issues are more of concern in recent years as multiple interim analyses (or even continuous monitoring) are increasingly used to improve the efficacy of basket and platform trials designs[12-14]. The other challenge is that most conventional phase II designs, such as Simon's two-stage design, assume a single primary endpoint. They are not equipped to handle co-primary endpoints (e.g., OR and PFS) that are of interest in some immunotherapy trials.

We propose a novel Bayesian adaptive design, namely time-to-event Bayesian optimal phase II (TOP) design, for immunotherapy trials. By adopting the Bayesian framework, the TOP design has several advantages over the conventional phase II designs, e.g., Simon's two-stage design. First, TOP is more efficient. It allows trialists to make real-time interim "go/no-go" decision when some patient's responses are pending, and perform more frequent interim monitoring to quickly terminate futile treatments. Such efficiency improvement translates into a shorter trial duration and fewer patients exposed to futile treatment. Secondly, TOP is more flexible. It accommodates various types of endpoints, including single endpoint (e.g., OR), co-primary efficacy endpoints (e.g., OR and PFS), and toxicity endpoints. In contrast to many Bayesian adaptive designs that often require intensive real-time computation and model fitting, the TOP design is transparent and easy to implement. The "go/no-go" decision rules of TOP can be tabulated and included in the trial protocol prior to the start of the trial. During the trial conduct, no complicated computation is needed, the "go/no-go" decision can be easily made by looking up the decision table. The explicit decision rules of the TOP design provide a valuable tool for guiding data monitoring deliberations and improves the transparency of the clinical trials process[15].

**Methods:**

**Three Trial examples**

*Example 1 (a binary efficacy endpoint).* A multicenter phase II study was conducted to evaluate the efficacy of nivolumab in patients with relapsed/refractory classic Hodgkin lymphoma after autologous hematopoietic cell transplantation treatment failure[16]. The primary endpoint was the OR rate (ORR), defined as the proportion of patients with an OR at any point assessed by the independent radiology review committee. The treatment is regarded as futile if the ORR is less than 20% and promising if the ORR is greater than 40%.

*Example 2 (co-primary efficacy endpoints).* A phase II trial was initiated at MD Anderson Cancer Center to treat recurrent isocitrate dehydrogenases (IDH) mutated glioma patients with hypermutator phenotype using nivolumab. The OR, scored using Immunotherapy Response Assessment in Neuro-Oncology (iRANO)[17], and PFS at 4 months (PFS4) are used as co-primary endpoints. The target ORR is 45% and target PFS4 rate is 30%. Along as one of the endpoints is above the target, the treatment is deemed promising.

*Example 3 (Jointly monitoring efficacy and toxicity).* Pembrolizumab is humanized antibody that targets the PD-1 immune checkpoint. In a phase II clinical trial, patients with relapsed/refractory classic hodgkin lymphoma were enrolled and treated[18]. The primary objective is to assess the safety and anti-tumor activity of pembrolizumab. The primary safety endpoint is the adverse events graded by National Cancer Institute Common Terminology Criteria for Adverse Events. The highest acceptable toxicity rate is 30%. The efficacy endpoint is the ORR by central review and the treatment is regarded as ineffective if ORR $\leq$ 15%.

**TOP design**

For ease of exposition, we first describe the TOP design using Example 1 with ORR as the primary endpoint, followed by a more general form of the TOP design that accommodates more complicated trial settings such as Examples 2 and 3. The TOP design consists of $R$ interim looks planned at a set of prespecified time points (e.g., after every 10 patients are enrolled). The TOP design is flexible and allows clinicians to choose the total number and timing of the interims based on clinical and logistic considerations. In general, more interim analyses lead to a more efficient design, enabling us to terminate futile treatments earlier and more frequently. Let $p$ denote the ORR of the treatment. The treatment is deemed ineffective if $p \leq \phi$, where $\phi$ is an ORR threshold prespecified by clinicians, e.g., $\phi = 20\%$. At each interim, the "go/no-go" decision is made based on the following Bayesian stopping criteria,

Stop the trial if $\Pr(p \leq \phi \mid Data) > C_n$; otherwise continue;

where the posterior probability $\Pr(p \leq \phi \mid Data)$ represents, given the interim data, how likely the true ORR is below the prespecified ORR threshold; and $C_n$ is a probability cutoff whose value depends on the interim sample size $n$ as described later. This Bayesian rule says that if the interim data suggest that the treatment is unlikely to reach the minimal efficacy requirement, then we stop the trial early for futility. The stopping cutoff $C_n$ is adaptive and depends on the interim sample size $n$, such that the stopping criteria are lenient at the beginning of the trial and become increasingly stricter when the trial proceeds. It reflects the practical consideration that at the beginning the trial, clinicians prefer a lenient efficacy requirement to avoid incorrectly stopping the trial too early (caused by spare data) so that more data can be collected to learn more on the characteristics of the agent; whereas toward the end of the trial, as more data accumulate, a stringent requirement is preferred so that the futile treatment can be stopped timely. The similar Bayesian stopping criteria were used by Zhou et al.[19] in the Bayesian optima phase II (BOP2) design, but the BOP2 assumes that the endpoint is quickly ascertainable.    Following Zhou et al.[19], we choose $C_n = 1 - C \left(\frac{n}{N}\right)^\gamma$, where $N$ is the total sample size, and $C$ and $\gamma$ are tuning parameters. To determine the values of $C$ and $\gamma$, we elicit from clinicians the null ORR that deems futile, the alternative ORR that deems promising, and the desirable type I error rate. We then calibrate the values of $C$ and $\gamma$ such that the type I error rate is controlled at a prespecified level and the statistical power is maximized. In the above Bayesian stopping criteria, posterior probability $\Pr(p \leq \phi \mid Data)$ is evaluated using a simple Bayesian statistical model, known as the beta-binomial model. The technical details of the model and the procedure of calibrating cutoff tuning parameters $C$ and $\gamma$ are provided in Supplementary data.

A major difficulty is that when the OR assessment window is long, by the time of making interim decisions, some patients have not completed their response assessment yet and their response outcomes are pending (or unknown). Table 1 shows interim data of 20 patients, where patients 1-11 have completed OR assessment with 3 responses and 8 nonresponses, and patients 12-20 only partially went through their assessment window and their response are unknown yet. Clearly, it is not appropriate to make a "go/no-go" decision with the interim data as 3/20 patients responded because the nine pending patients may turn out to be responses later, i.e., pending patients should not be counted equivalently as patients whose responses are known. On the other hand, we cannot ignore pending patients either because they do contain information (i.e., the patient has not progressed yet by the interim time). The TOP design overcomes this difficulty by assigning pending patients a partial credit according to his/her follow-up time using the notion of *effective sample size* (ESS), defined as ESS of a pending patient with a follow-up time $t = t/A$; where $A$ is the length of assessment window. ESS is the portion of follow-up that a pending patient has completed, reflecting the fact that the longer a patient has been followed, the more information that patient provides, and thus a larger credit should be assigned. For example, if the assessment takes 4 months (120 days), in Table 1, patients 12-20 are equivalent to an ESS of 0.71, 0.65, 0.55, 0.40, 0.27, 0.23, 0.08, 0.07, and 0.04, respectively. A patient whose OR status is known (e.g., patients 1-11) should receive a full credit of 1 as he/she provides full information. Thus, the total ESS (TESS) is $11 + 0.71 + 0.65 + \cdots + 0.04 = 14$. In other words, the interim data shown in Table 1 are equivalent to the "effective" data with 3/14 patients experienced OR. Based on this "effective" interim data, the Bayesian criteria (1) can be used to make a real-time "go/no-go" decision. In general, let $n_{obs}$ denote the number of patients whose response statuses are known, and $T_{total}$ be the sum of total follow-up times for all pending patients, the TESS is calculated as follows:

$$\text{TESS} = n_{obs} + T_{total}/A.$$

The statistical justification of TESS is provided in Supplementary data.

One appealing feature of the TOP design is that its decision rule can be pre-tabulated prior to the start of the trial. This is contrast to many Bayesian designs, which require complicated real-time model fitting and imputation. Given a maximum sample size of 40, Table 2 shows the "go/no-go" rule for trial Example 1 with interim analyses made after treating every 10 patients. To conduct the trial, we simply count the number of patients enrolled, the number of responses, the number of pending patients, and compute the TESS, and then use the table to make "go/no-go" decision. Suppose at an interim time, the observed data are given in Table 1. As the corresponding TESS = 14 and does not cross the stopping boundary of 15.4, according to the rule in Table 2, the decision is "go" (i.e., continue the trial). In principle, TOP supports continuous accrual. To avoid risky decisions caused by sparse data and stabilize the trial when the accrual is too fast, we impose an accrual suspension rule: At an interim decision time with sample size $n$, if the number of responses does not reach the boundary for "go" and more than $100n/N\%$ of the patients' response outcomes are pending, suspend the accrual to wait for more data to become available. This rule corresponds to "Suspend" in Table 2. A step-by-step example of conducting a phase II trial using the TOP design is provided Supplementary data.

Another attractive feature of TOP is that its decision table is invariant to the length of assessment window and accrual rate. This means that the same decision table can be used for trials with different assessment windows and accrual rates, which greatly simplifies preparation of the trial protocol. In addition, the TOP design's decision table contains, and thus can also be used for, the conventional case that response is quickly ascertainable, i.e., there is no pending patient.

We now describe a more general form of the TOP design that handles a simple binary endpoint (Example 1) and also more complicated outcomes (Examples 2 and 3) in a unified framework. The key is that although the endpoints in Examples 2 and 3 are different and more complicated than that in Example 1, they all can be represented using a variable $Y$ with $K$ distinct categories, statistically known as a multinomial random variable. For example, in Example 1, $Y$ has $K = 2$ categories (1= OR and 2= no OR); and in Example 2, $Y$ has $K = 4$ categories (i.e., 1 = (OR, PFS4), 2 = (OR, no PFS4), 3 = (no OR, PFS4), and 4 = (no OR, no PFS4)); and in Example 3, $Y$ has $K = 4$ categories (i.e., 1 = (OR, DLT), 2 = (OR, no DLT), 3 = (no OR, DLT), and 4 = (no OR, no DLT)). This unified endpoint $Y$ can be modelled using a Bayesian model, known as the Dirichlet-multinomial model. At each interim, the "go/no-go" decision for all examples can be made using similar Bayesian stopping criteria described previously. Specifically, for Example 2, let $p_{OR}$ and $p_{PFS6}$ denote the ORR and PFS4, and suppose that the treatment is deemed ineffective if $p_{OR} \leq \phi_{OR}$ and $p_{PFS4} \leq \phi_{PFS4}$, where $\phi_{OR}$ and $\phi_{PFS4}$ are ORR and PFS4 thresholds prespecified by clinicians, then interim "go/no-go" rule is defined as: Stop the trial if $\Pr(p_{OR} \leq \phi_{OR} \mid Data) > C_n$ and $\Pr(p_{PFS4} \leq \phi_{PFS4} \mid Data) > C_n$, otherwise continue. This rule says that we stop the trial if the interim data suggest that there is a high probability (i.e., $> C_n$) that the treatment is ineffective in both ORR and PFS4. Cutoff $C_n$ takes the same form as described previously. Although the TOP design does not require using the same cutoff for ORR and PFS4, we found that using the same cutoff simplifies the optimization of the design and still yields satisfactory operating characteristics. The "go/no-go" rule for Example 3 and the procedure of calibrating the rule to control type I error and maximize power are provided in Supplementary data.

Similar to the binary endpoint, the decision rules of the TOP design for Examples 2 and 3 can also be pre-tabulated prior to the start of the trial (see Table 3 and Table S1 in Supplementary data), making it simple to implement in practice. In addition, the same table can be used regardless of the length of assessment window and accrual rate, as the decision rule of TOP is invariant to these factors.

**Software**

To facilitate the application of the TOP design, we developed a user-friendly software (see Figure S1) that will be freely available at http://www.trialdesign.org. It allows users to generate the decision table as Table 1, evaluate operating characteristics of the design, and generate the trial design template for protocol preparation.

## 2. Numerical study

**Study design**

Simulation studies were carried out to examine the performance of the proposed TOP design, compared to Simon's two-stage design, and two Bayesian designs including the Bayesian futility monitoring design by Thall and Simon[20] (TS) and the BOP2 design[19]. Similar to Simon's design, the TS design assumes a single binary endpoint, but allows multiple interim analyses using a Bayesian criteria. BOP2 is more flexible than TS in that BOP2 not only allows multiple interim analyses, but also can handle various types of endpoints. All the three existing designs, i.e., Simon's two-stage design, TS and BOP2 designs, cannot handle delayed responses. To proceed, we adopt the aforementioned over-run approach for the Simon's two-stage design, and the suspending accrual approach for the TS and BOP2 designs. We considered nine scenarios. Scenarios 1-3 represent the case that ORR is the primary endpoint, similar to Example 1. The assessment window for OR is 4 months, and the treatment is deemed unacceptable if ORR $\leq 0.3$. The true ORR in scenarios 1-3 are 0.2, 0.3 and 0.5, corresponding to excessively ineffective, ineffective, and effective, respectively. Scenarios 4-6 consider the cases similar to Example 2, with 2-month ORR and PFS4 as the co-primary endpoints. The treatment is regarded as ineffective if the ORR $\leq 0.45$ and the PFS4 $\leq 0.30$. Scenarios 4-6 represent the cases of low ORR (=0.3) and low PFS4 (=0.3), low ORR (=0.4) but high PFS4 (=0.55), and high ORR (=0.65) and high PFS4 (=0.45), respectively. Scenarios 7-9 consider ORR and DLT rate as the co-primary endpoints, similar to Example 3. The treatment is deemed ineffective if ORR $\leq 0.3$, or overly toxic if the DLT rate $\geq 0.3$. The assessment window for ORR and toxicity are 4 and 2 months, respectively. Scenarios 4-6 correspond to low ORR (=0.3) and high toxicity (=0.3), high ORR (=0.45) and high toxicity (=0.4), and high ORR (=0.5) and low toxicity (=0.18), respectively. The details on the statistical joint distributions of the two endpoints in scenarios 4-9 are provided in Supplementary Data. Since the Simon's two-stage design and the TS design can handle only a single endpoint, in scenarios 4-9, ORR is used as the endpoint of these two designs. Under each scenario, the maximum sample size is determined based on the Simon's two-stage design. Specifically, in scenarios 1-3 and 7-9, 46 patients are required by the Simon's two-stage design to test the null hypothesis that the ORR$\leq 0.3$ versus the alternative hypothesis that the ORR$\geq 0.5$ with a type I error rate of 0.1 and a type II error rate of 0.1. In the Simon's two-stage design, one interim analysis is performed after 22 patients are enrolled. For TOP, TS and BOP2, a total of 3 interim analyses are performed after each 12 patients are treated. In scenarios 4-6, 40 patients are required by the Simon's two-stage design to test the null hypothesis that the ORR$\leq 0.45$ versus the alternative hypothesis that the ORR$\geq 0.65$ with a type I error rate of 0.1 and a type II error rate of 0.15. In the Simon's two-stage design, one interim analysis is performed after 20 patients are enrolled. For TOP, TS and BOP2, a total of 3 interim analyses are performed after each 10 patients are treated. In all scenarios, the accrual rate is 2 patients/month. The time-to-event outcomes (e.g., OR, PFS4 and DLT) are sampled from a Weibull distribution with 50% of the events occurred in the second half of the response-specific assessment window. See the Supplementary Data for details on the simulation configurations.

**Performance metrics**

Four performance metrics are used to characterize the operating characteristics of the designs based on 10,000 simulated trials.

1. Type I error (or false positive) rate, measuring the likelihood that a design incorrectly claims an ineffective treatment as effective

2. Power, measuring the likelihood that a design detects an effective treatment

3. Expected sample size

4. Average trial duration

**Results**

*A binary efficacy endpoint (scenarios 1-3)*

Figures 2 reports the simulation results. When the treatment is ineffective (i.e., scenarios 1 and 2), all the methods successfully control type I error rate < 10%. The TOP design has a substantailly smaller expected sample size and shorter trial duration than the Simon's design, demonstrating the efficiency gain of using multiple interim looks. The TS design has the smallest sample size and shortest trial duration, but it has lower power than the TOP design (77.4% vs 90.7%) when the treatment is effective (i.e., scenario 3). TOP and BOP2 have similar type I error rates, powers and expected sample sizes, but the average trial durations of TOP are about 4~10 months shorter than those of BOP2. This is because TOP allows for real-time decision making, while BOP2 requires suspending the accrual to obtain fully observed data to make decisions.

*ORR and PFS as co-primary endpoints (scenarios 4-6)*

When the treatment is ineffective (i.e., scenario 4), all the designs are capable to control the type I error rate at the level of 10%. Compared to the Simon's design, TOP has a smaller expected sample size and shorter trial duration. When the treatment is effective (i.e., scenarios 5 and 6), TOP yields substantailly higher power than the Simon's two-stage design and TS design, demonstrating the advantage of considering co-primary endpoints. In particular, in scenario 5, where PFS is favorable, the power for TOP is 87.2%, whereas that for Simon's design and TS design is merely 2.8% and 2.6%. BOP2 considers co-primary endpoint and has comparable performance as TOP, but it leads to longer trial duration because of suspending accrual.

*Monitoring ORR and toxicity jointly (scenarios 7-9)*

Under the null case that the treatment is ineffective and toxic (scenario 7), all designs control the type I error rate at 10%. In scenario 8, the treatment has a high ORR but is unacceptable due to toxicity. The Simon's and TS designs failed to recognize that and claim that the treatment is promising 62.9% and 76.6% of the time because they ignore toxicity. In contrast, TOP only has 4.9% chance to incorrectly claim that the treatment is promising, and terminated trial early with a smaller number of patients exposed to the toxic treatment. Scenario 9 presents the ideal case that the treatment is safe and effective. Simon's two-stage design, TOP and BOP2 have similar power and outperform the TS design. Again, TOP and BOP2 have similar performance in terms of type I error rate, power and sample size, but TOP has a shorter trial duration.

*Sensitivity analysis*

Extensive numerical studies are conducted to evaluate the robustness of the TOP design. The results show that the performance of TOP is robust to the length of assessment window, patient accrual rate, degree of delay, time-to-response distribution, and the calculation scheme of TESS (see the Supplementary Data).

## 3. Discussion

Immunotherapy trials are challenging because of potentially late-onset responses and complicated endpoints. We have proposed a flexible Bayesian phase II design to address these challenges. Our TOP design allows real-time "go/no-go" decision making when some patients' outcomes are pending, thereby shortening the trial duration and avoiding treating many patients at an ineffective treatment. TOP is flexible and can handle various types of endpoints under a unified framework. It is also efficient, allows any arbitrary number of interim looks, and maximizes the power with well controlled type I error rate by incorporating all available trial information in decision making. The decision boundaries of TOP can be pre-tabulated and included in the protocol before the trial starts, making it transparent and particularly easy to implement in practice.

This article focuses on single-arm phase II trials. The TOP design can be extended to randomized two or multiple arms phase II trials, where we compare the efficacy (e.g., ORR) of the experimental arm to that of the control arm, rather a fixed historical control (e.g., 30%). The similar Bayesian "go/no-go" criteria, e.g., stop the trial if $\Pr\left( ORR_{exp} > ORR_{ctl} \mid Data \right) > C_n$, can be used to make interim decisions, where $ORR_{exp}$ and $ORR_{ctl}$ are ORRs for experimental and control arms, respectively.

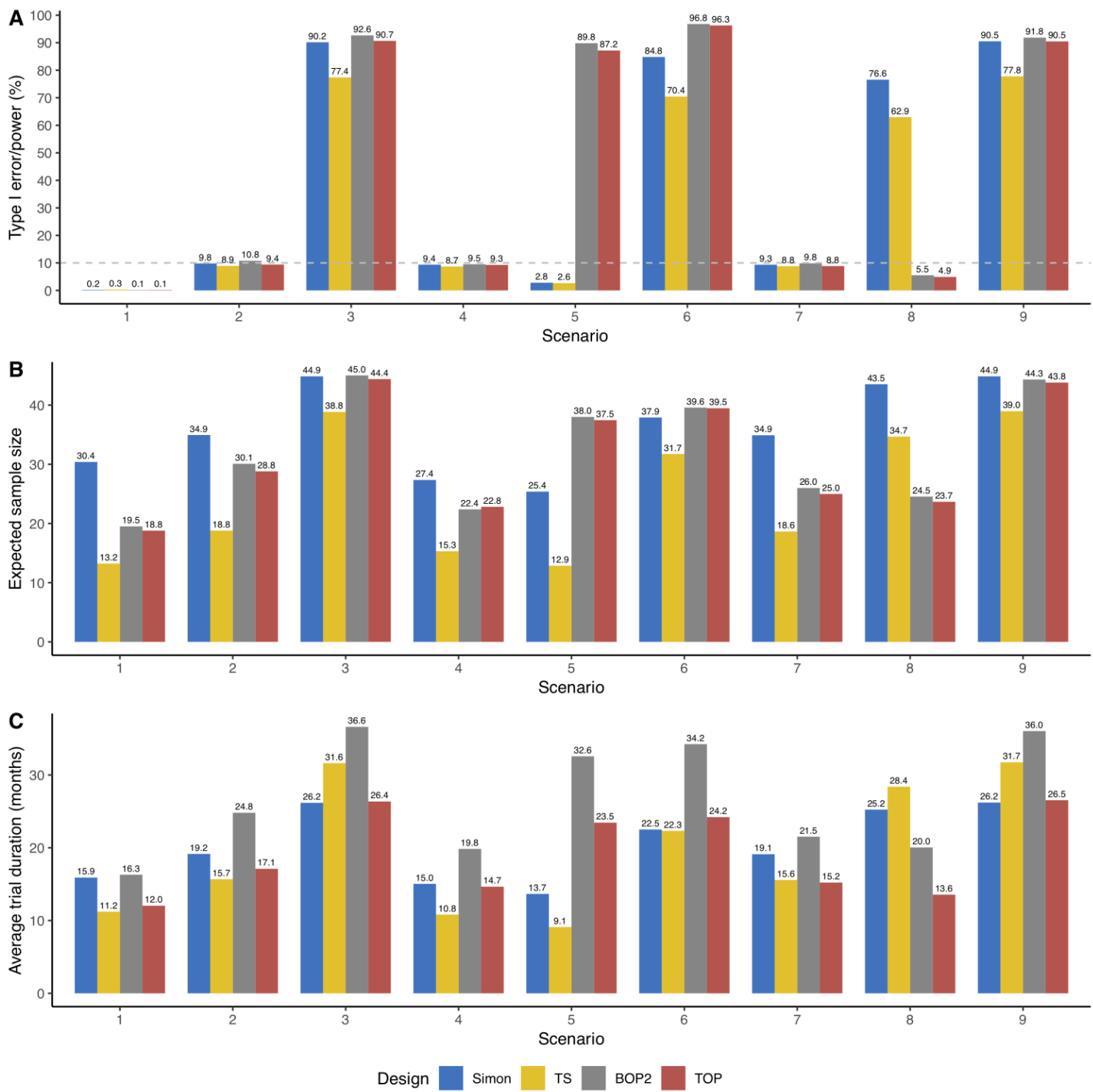

**Figure 2**. Simulation results for four phase II trial designs, where Simon is the Simon's two-stage design[3], TS is the Bayesian futility monitoring design proposed by Thall and Simon[18], BOP2 is the Bayesian optimal phase II trial design[17], and TOP is the proposed time-to-event Bayesian optimal phase II trial design. Scenarios 1-3 represent the case that ORR is the primary endpoint, similar to Example 1. Scenarios 4-6 consider the case with 2-month ORR and the 4-month PFS (PFS4) as the co-primary endpoints, similar to Example 2. Scenarios 7-9 consider ORR and DLT rate as the co-primary endpoints, similar to Example 3. In panel A, Scenarios 1, 2, 4, and 7 show the type I error rate, and the remaining scenarios show the statistical power.

**Table 1**. An example of interim data from 20 patients with the response assessment window of 120 days, of whom the first 10 have the outcomes observed and the last 10 have pending outcomes. The total ESS (TESS) of the 20 patients is 14.

| Patient id | Assessment finished? If no, then follow-up time. | Response? | ESS | Patient id | Assessment finished? If no, then follow-up time. | Response? | ESS |
|---|---|---|---|---|---|---|---|
| 1 | Yes | No | 1 | 11 | Yes | Yes | 1 |
| 2 | Yes | Yes | 1 | 12 | No, 85 days | Pending | 0.71 |
| 3 | Yes | No | 1 | 13 | No, 78 days | Pending | 0.65 |
| 4 | Yes | Yes | 1 | 14 | No, 66 days | Pending | 0.55 |
| 5 | Yes | No | 1 | 15 | No, 48 days | Pending | 0.40 |
| 6 | Yes | No | 1 | 16 | No, 32 days | Pending | 0.27 |
| 7 | Yes | No | 1 | 17 | No, 28 days | Pending | 0.23 |
| 8 | Yes | No | 1 | 18 | No, 10 days | Pending | 0.08 |
| 9 | Yes | No | 1 | 19 | No, 8 days | Pending | 0.07 |
| 10 | Yes | No | 1 | 20 | No, 5 days | Pending | 0.04 |

ESS: effective sample size, defined as the patient's follow-up time divided by the assessment window (e.g., 120 days). For patients who have completed the response assessment, ESS=1.

**Table 2**. Go/no-go rule for TOP with a single binary endpoint using the setting of Example 1. The maximum sample size is 40 and interim analyses are made after treating every 10 patients.

| # Patients | # Responses | # Pending patients | Action | # Patients | # Responses | # Pending patients | Action |
|---|---|---|---|---|---|---|---|
| 10 | $\leq 1$ | $\geq 2$ | Suspend | 30 | $\leq 2$ | $\leq 21$ | No Go |
| 10 | 0 | $\leq 1$ | No go | 30 | 3 | $\leq 21$ | Go if TESS < 11.44 |
| 10 | 1 | $\leq 1$ | Go if TESS < 8.27 | 30 | 4 | $\leq 21$ | Go if TESS < 15.98 |
| 10 | $\geq 2$ | $\leq 8$ | Go | 30 | 5 | $\leq 21$ | Go if TESS < 20.57 |
| 20 | $\leq 3$ | $\geq 10$ | Suspend | 30 | 6 | $\leq 21$ | Go if TESS < 25.21 |
| 20 | $\leq 1$ | $\leq 9$ | No go | 30 | 7 | $\leq 21$ | Go if TESS < 29.88 |
| 20 | 2 | $\leq 9$ | Go if TESS < 10.15 | 30 | $\geq 8$ | $\leq 22$ | Go |
| 20 | 3 | $\leq 9$ | Go if TESS < 15.40 | 40 | $\leq 39$ | $\geq 1$ | Suspend |
| 20 | $\geq 4$ | $\leq 16$ | Go | 40 | $\leq 11$ | 0 | No go |
| 30 | $\leq 7$ | $\geq 22$ | Suspend | 40 | $\geq 12$ | 0 | Go |

"Go" means continue the trial to the next interim analysis; "No go" means stop the trial early; "Suspend" means suspending the accrual to wait for more data available. "TESS" is the total effective sample size. Decision rule is derived such that the type I error rate is controlled at the nominal level of 0.1 under the null hypothesis $ORR = 0.20$ and the statistical power is maximized under the alternative hypothesis $ORR = 0.40$.

**Table 3**. Go/no-go rule for TOP with ORR and PFS at 4 months (PFS4) as the co-primary endpoints using the setting of Example 2. The maximum sample size is 45 and interim analyses are made after treating every 15 patients. "No go" decision should be made if the treatment is futile in terms of both ORR and PFS4; otherwise, "go" decision should be made and the trial continues to the next interim analysis.

| Endpoint 1 (ORR) | | | |
|---|---|---|---|
| # Patients | # Responses | # Pending patients | Futile? |
| 15 | $\leq 6$ | $\geq 5$ | Suspend |
| 15 | $\leq 4$ | $\leq 4$ | Yes |
| 15 | 5 | $\leq 8$ | Yes if TESS $\geq 10.65$ |
| 15 | 6 | $\leq 8$ | Yes if TESS $\geq 12.83$ |
| 15 | $\geq 7$ | $\leq 8$ | No |
| 30 | $\leq 15$ | $\geq 20$ | Suspend |
| 30 | $\leq 5$ | $\leq 19$ | Yes |
| 30 | 6 | $\leq 19$ | Yes if TESS $\geq 10.60$ |
| 30 | 7 | $\leq 19$ | Yes if TESS $\geq 12.59$ |
| 30 | 8 | $\leq 19$ | Yes if TESS $\geq 14.59$ |
| 30 | 9 | $\leq 19$ | Yes if TESS $\geq 16.61$ |
| 30 | 10 | $\leq 19$ | Yes if TESS $\geq 18.63$ |
| 30 | 11 | $\leq 19$ | Yes if TESS $\geq 20.67$ |
| 30 | 12 | $\leq 19$ | Yes if TESS $\geq 22.72$ |
| 30 | 13 | $\leq 19$ | Yes if TESS $\geq 24.77$ |
| 30 | 14 | $\leq 19$ | Yes if TESS $\geq 26.83$ |
| 30 | 15 | $\leq 19$ | Yes if TESS $\geq 28.89$ |
| 30 | $\geq 16$ | $\leq 14$ | No |
| 45 | $\leq 44$ | $\geq 1$ | Suspend |
| 45 | $\leq 25$ | 0 | Yes |
| 45 | $\geq 26$ | 0 | No |
| Endpoint 2 (4-month PFS) | | | |
| # Patients | # Progression free at 4 months | # Pending patients | Futile? |
| 15 | $\leq 4$ | $\geq 5$ | Suspend |
| 15 | $\leq 3$ | $\leq 4$ | Yes |
| 15 | 4 | $\leq 4$ | Yes if TESS $\geq 12.32$ |
| 15 | $\geq 5$ | $\leq 10$ | No |
| 30 | $\leq 11$ | $\geq 20$ | Suspend |
| 30 | $\leq 4$ | $\leq 19$ | Yes |

| 30 | 5 | ≤ 19 | Yes if TESS ≥ 12.20 |
|----|----|------|---------------------|
| 30 | 6 | ≤ 19 | Yes if TESS ≥ 15.10 |
| 30 | 7 | ≤ 19 | Yes if TESS ≥ 18.03 |
| 30 | 8 | ≤ 19 | Yes if TESS ≥ 20.99 |
| 30 | 9 | ≤ 19 | Yes if TESS ≥ 23.97 |
| 30 | 10 | ≤ 19 | Yes if TESS ≥ 26.97 |
| 30 | 11 | ≤ 19 | Yes if TESS ≥ 29.99 |
| 30 | ≥ 12 | ≤ 18 | No |
| 45 | ≤ 44 | ≥ 1 | Suspend |
| 45 | ≤ 18 | 0 | Yes |
| 45 | ≥ 19 | 0 | No |

"Suspend" means suspending the accrual to wait for more data available. "TESS" is the total effective sample size. Decision rule is derived such that the type I error rate is controlled at the nominal level of 0.1 under the null hypothesis $\mathrm{ORR} = 0.45$ and $\mathrm{PFS4} = 0.30$ and the statistical power is maximized under the alternative hypothesis $\mathrm{ORR} = 0.65$ and $\mathrm{PFS4} = 0.45$.